\documentclass[twocolumn]{jpsj2} %% two-column layout
\title{Evolution of Heterogeneous Antiferromagnetic State in URu$_2$Si$_2$: Study of Hydrostatic-Pressure, Uniaxial-Stress and Rh-Dope Effects}

\author{Makoto \textsc{Yokoyama}$^{1}$\thanks{E-mail address: makotti@mx.ibaraki.ac.jp} and Hiroshi \textsc{Amitsuka}$^{2}$}

\inst{
$^{1}$Faculty of Science, Ibaraki University, Mito 310-8512\\
$^{2}$Graduate School of Science, Hokkaido University, Sapporo 060-0810\\
}

\def\runauthor{M.\ \textsc{Yokoyama} and H.\ \textsc{Amitsuka}}

\abst{
We have investigated the nature of the competition between hidden order and antiferromagnetic (AF) order in URu$_2$Si$_2$ by performing the neutron scattering experiments under hydrostatic-pressure $P$, uniaxial-stress $\sigma$, and Rh-substitution conditions. Hidden order observed at ambient pressure in pure URu$_2$Si$_2$ is found to be replaced by the AF order by applying $P$, $\sigma$ along the tetragonal basal plane, and by doping Rh. We discuss these experimental results on the basis of the crystalline strain calculations, and suggest that this phase transition is generated by the 0.1\% increase of the tetragonal $c/a$ ratio. We have also found that the magnetic excitation observed in the hidden order phase vanishes in the AF phase. We show that this variation can be qualitatively explained by assuming the hidden order parameter to be quadrupole.  
}

\kword{heavy-electron system, URu$_2$Si$_2$, hidden order, inhomogeneous magnetism, neutron scattering}

\begin{document}
\maketitle
\section{Introduction}
The ternary tetragonal compound URu$_2$Si$_2$ is a well known heavy-electron system that shows a peculiar ordered phase below $T_{\rm o}=17.5\ {\rm K}$ and an unconventional superconductivity below $T_{\rm c}\sim 1.2\ {\rm K}$.\cite{rf:Palstra85,rf:Schlabitz86,rf:Maple86} The phase transition at $T_{\rm o}$ is associated with clear anomalies in bulk properties including specific heat, magnetic susceptibility and electrical resistivity, whose shapes indicate that the gap is formed in parts of the Fermi surfaces of the heavy quasi-particles below $T_{\rm o}$. The elastic neutron scattering experiments\cite{rf:Broholm87,rf:Broholm91,rf:Mason90,rf:Mason95,rf:Fak96,rf:Honma99} revealed that the type-I antiferromagnetic (AF) order develops below $\sim T_{\rm o}$. However, the magnitude of the AF moment $\mu_{\rm o}$ is found to be extremely small ($\sim 0.03\ \mu_{\rm B}/{\rm U}$), which is hard to account for the large bulk anomalies at $T_{\rm o}$ such as the specific-heat jump ($\Delta C/T_{\rm o}\sim 300\ {\rm mJ/K^2\ mol}$). Furthermore, the internal field associated with $\mu_{\rm o}$ is never observed in the $^{29}$Si-NMR experiments.\cite{rf:Kohara86} The inconsistency among these experiments has been puzzling many researchers for about 20 years, i.e., whether the intrinsic order parameter is the weak AF moment or some unidentified ``hidden" degree of freedom. So far various ideas for the order parameter have been proposed: the AF moment with highly reduced $g$ values,\cite{rf:Nieuwenhuys87,rf:Sikkema96,rf:Okuno98,rf:Yamagami2000,rf:Bernhoeft2003} quadrupole order,\cite{rf:Miyako91,rf:Ami94,rf:Santini94,rf:Ohkawa99,rf:Tsuruta2000} octupole order,\cite{rf:Kiss2004,rf:Hanzawa2005} valence transition,\cite{rf:Barzykin95} uranium dimers,\cite{rf:Kasuya97} unconventional spin density wave,\cite{rf:Ikeda98,rf:Virosztek2002} charge current order,\cite{rf:Chandra2001} and helicity order.\cite{rf:Varma2006}

We expect that the investigations on the relationship between the weak AF moment and the bulk properties may provide crucial keys to resolve this issue. It is remarkable that the reported $\mu_{\rm o}(T)$ curves differ largely from each other,\cite{rf:Broholm87,rf:Broholm91,rf:Mason90,rf:Mason95,rf:Fak96,rf:Honma99} in contrast to the mean-field-like anomalies of bulk properties at $T_{\rm o}$. In addition, the neutron scattering and the specific heat measurements for the samples prepared under annealing and as-grown conditions revealed that $\mu_{\rm o}(T)$ shows more prominent sample-quality dependence than the specific heat.\cite{rf:Fak96} This difference implies that the evolution of the weak AF moment depends highly on the crystal conditions. On the other hand, it is expected that many of the proposed order parameters involve a magnetic instability such that the large-moment AF order may be generated by the crystal distortions. It is therefore interesting to investigate the role of the crystal strains on the AF moment using the microscopic probes. In this paper, we present the results of our recent neutron scattering experiments performed on URu$_2$Si$_2$ under hydrostatic pressure, uniaxial stress, and on U(Ru$_{1-x}$Rh$_x$)$_2$Si$_2$.

\section{Elastic Neutron Scattering under Hydrostatic Pressure and Uniaxial Stress}
We have investigated the effects of hydrostatic pressure\cite{rf:Ami99,rf:Ami2000} and uniaxial stress\cite{rf:Yoko2002,rf:Yoko2005} on the AF moment by performing the elastic neutron scattering experiments. Hydrostatic pressure $P$ was applied by means of a piston-cylinder device,\cite{rf:Onodera87} and uniaxial stress $\sigma$ was applied along the [100], [110], and [001] directions using a clamp-type pressure cell. The details of the experimental conditions were described elsewhere\cite{rf:Ami99,rf:Ami2000,rf:Yoko2002,rf:Yoko2005}. In Fig.\ 1, we plot the $P$ and $\sigma$ variations of the volume-averaged AF moment $\mu_{\rm o}$ at 1.5 K. The $\mu_{\rm o}$ values at ambient pressure were estimated to be $\sim 0.02\ \mu_{\rm B}/{\rm U}$ for all experiments, which are consistent with the previous ones.\cite{rf:Broholm87,rf:Broholm91,rf:Mason90,rf:Mason95,rf:Fak96,rf:Honma99} By applying $P$, $\mu_{\rm o}$ continuously develops to 0.25 $\mu_{\rm B}/{\rm U}$ ($P=1.0\ {\rm GPa}$), and then shows an abrupt increase to $0.4\ \mu_{\rm B}/{\rm U}$ at $\sim 1.5\ {\rm GPa}$. On the other hand, a significant anisotropy is found in the $\sigma$ variations of $\mu_{\rm o}$: $\mu_{\rm o}$ is enhanced to $0.25\ \mu_{\rm B}/{\rm U}$ ($\sigma=0.4\ {\rm GPa}$) with applying $\sigma$ along the [100] and [110] directions, while it shows only a slight increase to $0.028\ \mu_{\rm B}/{\rm U}$ ($\sigma=0.46\ {\rm GPa}$) for $\sigma\parallel [001]$. The $\mu_{\rm o}(\sigma)$ curve for $\sigma\parallel [100]$ coincides fairly with that for $\sigma\parallel [110]$, indicating that the development of $\mu_{\rm o}$ is isotropic with respect to the compression in the tetragonal basal plane. Interestingly, the increasing rate of $\mu_{\rm o}$ for $\sigma\parallel [100]$ and [110] ($\partial \mu_{\rm o}/\partial \sigma \sim 1.0\ \mu_{\rm B}/{\rm GPa}$) is four-fold larger than that for $P$ ($\partial \mu_{\rm o}/\partial \sigma \sim 0.25\ \mu_{\rm B}/{\rm GPa}$). The application of $\sigma$ is thus considered to enhance the $\mu_{\rm o}$ value more effectively than $P$. In contrast to the sensitive variations of the $\mu_{\rm o}$ values under $P$ and $\sigma$, the onset of $\mu_{\rm o}$ shows only a slight increase with applying $P$, $\sigma\parallel [100]$ and $\sigma \parallel[110]$: the rates of increase are roughly estimated to be $\sim 1.0\ {\rm K}/{\rm GPa}$.
\begin{figure}[tbp]
\begin{center}
\includegraphics[keepaspectratio,width=0.5\textwidth]{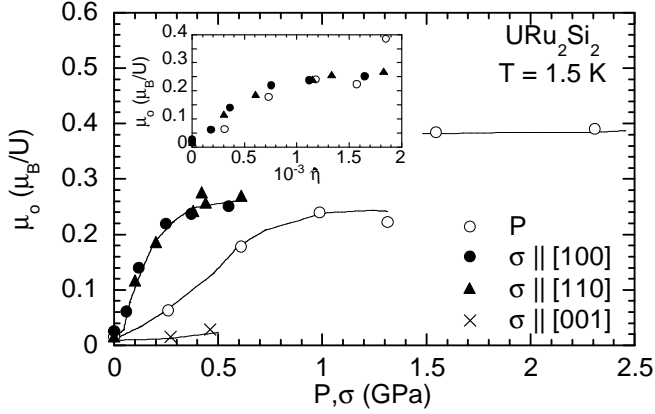}
\end{center}
  \caption{Hydrostatic-pressure ($P$) and uniaxial-stress ($\sigma$) variations of the volume-averaged AF moment $\mu_{\rm o}$ at 1.5 K in URu$_2$Si$_2$. The inset shows the $\mu_{\rm o}$ data obtained under $P$, $\sigma \parallel [100]$ and $\sigma \parallel [110]$, plotted as a function of $\hat{\eta }$ ($\equiv (\eta - \eta_0)/\eta_0;\ \eta=c/a$). The lines are guides to the eye.}
\end{figure}

We have found that the uniaxial stress applied along the tetragonal basal plane brings about similar characteristics of the AF order, magnitude of $\mu_{\rm o}$ and its $\sigma$ dependence, to those given by hydrostatic pressure. This implies that there is an implicit and common parameter leading to an equivalent effect in the different types of compression. Our neutron scattering experiments revealed the following features for the enhancement of $\mu_{\rm o}$: (i) $\mu_{\rm o}$ increases by applying $P$ and $\sigma\bot [001]$, while it is nearly constant for $\sigma\parallel [001]$, (ii) the application of $\sigma$ enhances $\mu_{\rm o}$ more effectively than $P$, and (iii) the $\mu_{\rm o}(\sigma)$ curve is isotropic for the compression in the tetragonal basal plane. We suggest that these features can be explained by considering the implicit parameter to be the tetragonal $c/a$ ratio ($\equiv \eta$). In general, the crystalline strains induced in the tetragonal structure can be categorized into two types of the symmetry-invariant strains (volume and $c/a$ ratio) and four types of the symmetry-breaking strains ($x^2-y^2$, $xy$, $yz$ and $zx$). It is unlikely that the features (i) and (iii) are attributed to the couplings between $\mu_{\rm o}$ and the symmetry-breaking strains. On the other hand, the features (ii) and (iii) imply that the elongation of the $c/a$ ratio induces the AF state more sensitively than the volume compression. To check this, we calculated the $P$ and $\sigma$ variations of $\eta$ using the values of the elastic constants estimated from the ultrasonic sound velocity measurements.\cite{rf:Wolf94} In the inset of Fig.\ 1, we plot the $\mu_{\rm o}$ data obtained under $P$, $\sigma\parallel [100]$ and $\sigma\parallel [110]$ as a function of $\hat{\eta}$ $(\equiv (\eta-\eta_0)/\eta_0)$. It is found that they are well scaled by $\hat{\eta}$, and the critical value $\hat{\eta}_{\rm c}$ to enhance $\mu_{\rm o}$ is roughly $0.1\%$.

\section{The Nature of the Inhomogeneous AF State}
Recently, the high-pressure $^{29}$Si-NMR experiments revealed that the system is spatially separated into two different ordered regions below $T_{\rm o}$ under $P$: one is AF with a large moment and the other is non magnetic (or very weakly magnetic).\cite{rf:Matsuda2001,rf:Matsuda2003} The volume fraction of the AF state $V_{\rm AF}$ is found to continuously increase by applying $P$, while the magnitude of the internal field at the Si sites is nearly independent of $P$. Since the $P$ variations of $V_{\rm AF}$ are roughly in proportional to $\mu_{\rm o}^2(P)$, the enhancement of $\mu_{\rm o}$ observed in our neutron scattering experiments under $P$ is considered to be attributed to the increase of $V_{\rm AF}$, not of the local AF moment. In addition, $V_{\rm AF}$ at ambient pressure is estimated to be about 1\% if the $^{29}$Si-NMR data under $P$ are simply extrapolated to ambient pressure. It is suggested that this is the true nature of the ``weak" AF moment. The remaining $\sim 99\%$ is thus considered to be occupied by unidentified hidden order (HO), which is responsible to the large bulk anomalies at $T_{\rm o}$.

The development of the AF phase in the high-$P$ region is also observed in recent thermal expansion,\cite{rf:Motoyama2003} $\mu$SR,\cite{rf:Ami2003,rf:Amato2004} and neutron scattering \cite{rf:Bourdarot2004} measurements under $P$. Figure 2 shows the $P-T$ phase diagram obtained from our $\mu$SR experiments under $P$.\cite{rf:Ami2003,rf:Amato2004} We have observed that the AF volume fraction suddenly increases above $\sim 0.4\ {\rm GPa}$. The temperature variations of $V_{\rm AF}$ in the pressure range of 0.4--1 GPa show steep changes at $T_{\rm M}$ ($<T_{\rm o}$), strongly suggesting that the transition from HO to AF is of the 1st order. The thermal expansion \cite{rf:Motoyama2003} and neutron scattering \cite{rf:Bourdarot2004} experiments have also pointed out the existence of the 1st-order phase boundary in the $P-T$ phase diagram, although its position depends strongly on the sample and pressure conditions.
\begin{figure}[bp]
\begin{center}
\includegraphics[keepaspectratio,width=0.5\textwidth]{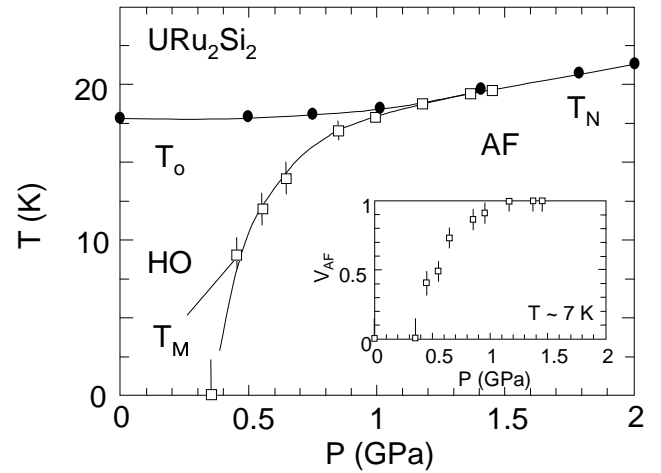}
\end{center}
  \caption{$P-T$ phase diagram of URu$_2$Si$_2$, estimated from the $\mu$SR experiments under $P$. $T_{\rm M}$ indicates the onset of $V_{\rm AF}$. $P$ variations of $V_{\rm AF}$ at $\sim 7\ {\rm K}$ are shown in the inset. The lines are guides to the eye.}
\end{figure}

The discontinuous increase of $V_{\rm AF}$ observed in the high-$P$ region contrasts remarkably with the $\mu_{\rm o}(P)$ curve obtained from our neutron scattering experiments: $\mu_{\rm o}$ is continuously enhanced with increasing $P$. We suggest that this difference may be ascribed to the distribution of the $\eta$ value in the sample due to some imperfection of the crystal. The width of the distribution is expected to be of the order of $\sim 0.01\%$, which will be hard to detect and analyze using usual microscopic probes. As is argued in Section 2, we consider that the AF phase is generated by satisfying the condition $\hat{\eta} >\hat{\eta}_c$. At ambient pressure, majority of the sample is occupied by HO below $T_{\rm o}$ because the mean value of $\hat{\eta}$ should be smaller than $\hat{\eta}_{\rm c}$. However, the $\hat{\eta}$ value in the small fragmentary regions may exceeds $\hat{\eta}_{\rm c}$ for the distribution of $\hat{\eta}$. The AF order takes place in such regions, being detected in the neutron scattering experiments as a very small moment on volume averaging. On the other hand, the mean value of $\hat{\eta}$ increases with increasing $P$. This is accompanied by the increase of the regions satisfying the condition $\hat{\eta} >\hat{\eta}_c$, observed as the development of the inhomogeneous AF phase. In such a situation, the $V_{\rm AF}(P)$ function is governed by the width and the shape of the $\hat{\eta}$'s distribution: $V_{\rm AF}$ is steeply enhanced at the critical pressure if the width of the distribution is small, while it continuously increases if the width is large. We expect that this effect causes the difference of $V_{\rm AF}(P)$ (and $\mu_{\rm o}(P)$) among the experiments. We note that the scaling behavior of our $\mu_{\rm o}(P)$ and $\mu_{\rm o}(\sigma)$ data is also consistent with the above consideration, since the samples used in our neutron scattering experiments under $P$ and $\sigma$ were cut out from the same ingot. In addition, we have very recently observed in the high-$P$ neutron scattering experiments using a good-quality sample \cite{rf:Ami2006} that $\mu_{\rm o}$ is enhanced very sharply at $P_{\rm c}\sim 0.7\ {\rm GPa}$. Above $P_{\rm c}$, the $\mu_{\rm o}$ value reaches $\sim 0.4\ \mu_{\rm B}/{\rm U}$, which corresponds to that observed in our previous neutron scattering experiments above $\sim 1.5\ {\rm GPa}$ (see Fig.\ 1). We thus consider that the discontinuous change of the $\mu_{\rm o}$ value at $\sim 1.5\ {\rm GPa}$ in our previous neutron scattering experiments is also ascribed to the effect of the AF volume fraction.

\section{Neutron Scattering Experiments on Rh-doped System}
We have suggested that the $0.1\%$ increase of the $\hat{\eta}$ value generates the HO-to-AF phase transition. It is expected that the $\hat{\eta}$ values can be tuned not only by applying the compressions but also by alloying. In the Rh-substitution system U(Ru$_{1-x}$Rh$_x$)$_2$Si$_2$, HO is suppressed at $x\sim 0.04$, and the non-magnetic heavy-electron state appears in the range between $x\sim 0.04$ and 0.1.\cite{rf:Ami88,rf:Burlet92} At the same time, the Rh substitution enhances the $\hat{\eta}$ value at the rate $\partial \hat{\eta}/\partial x \sim 7 \times 10^{-2}$, indicating that the 0.1\% increase of $\hat{\eta}$ is achieved at $x\sim 0.02$. We thus expect that the inhomogeneous AF state is induced by a small amount of Rh-doping. To verify this possibility, we have investigated the microscopic properties of U(Ru$_{1-x}$Rh$_x$)$_2$Si$_2$ in the small $x$ range between $x=0$ and 0.04, by performing the elastic and inelastic neutron scattering experiments.\cite{rf:Yoko2004}

Figure 3 shows temperature variations of the AF Bragg-peak intensities ($\propto \mu_{\rm o}^2$), obtained from the elastic neutron scattering experiments. For $x\le 0.015$, we have observed that $I(T)$ starts increasing below $\sim T_{\rm o}$, and shows a slow development with decreasing temperature. The $\mu_{\rm o}$ values still stay in the same order as that at $x=0$ ($\sim 0.02\ \mu_{\rm B}/{\rm U}$). On the other hand, $I(T)$ for $x=0.02$ slightly develops below $\sim 13.7\ {\rm K}$ ($\sim T_{\rm o}$), and then abruptly increases at $\sim 8.3\ {\rm K}$ ($=T_{\rm M}$). The $\mu_{\rm o}$ values at 1.4 K is estimated to be $0.24(1)\ \mu_{\rm B}/{\rm U}$. Similar $I(T)$ curves are also obtained for $x=0.025$ and 0.03, where the interval between $T_{\rm o}$ and $T_{\rm M}$ become narrower with increasing $x$. For $x=0.04$, no anomaly is found at least in the investigated temperature range between 1.4 K to 40 K. Except the suppression of both HO and AF phases at $x=0.04$, the obtained $x-T$ phase diagram compares reasonably well with the $P-T$ phase diagram for pure URu$_2$Si$_2$, suggesting that the Rh substitution brings about similar effects to those generated by applying $P$ for pure URu$_2$Si$_2$.\cite{rf:Bourdarot2004,rf:Yoko2004}
\begin{figure}[tbp]
\begin{center}
\includegraphics[keepaspectratio,width=0.5\textwidth]{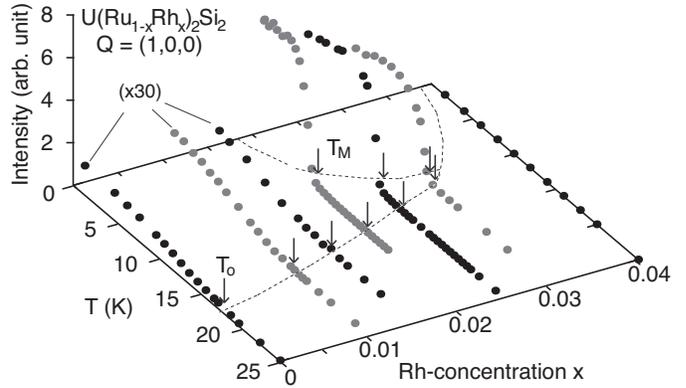}
\end{center}
  \caption{Temperature variations of the integrated intensities for the (100) magnetic reflections of U(Ru$_{1-x}$Rh$_x$)$_2$Si$_2$ ($x\le 0.04$). $T_{\rm o}$ and $T_{\rm M}$ are indicated by the arrows. Note that the intensities for $x=0$, 0.01 and 0.015 are 30-fold enlarged. The lines are guides to the eye.}
\end{figure}

Displayed in Fig.\ 4 is the temperature variations of the magnetic excitation spectra at the magnetic Brillouin zone center $Q=(1,0,0)$ for $x=0.02$, obtained from the inelastic neutron scattering experiments. The sharp peaks at the energy transfer $\hbar\omega =0$ mainly arise from the magnetic Bragg scattering and the incoherent nuclear scattering. A broad magnetic-excitation peak is found to develop below $T_{\rm o}=13.7\ {\rm K}$. The peak position is estimated to be zero within the experimental accuracy, which is significantly reduced from that for pure URu$_2$Si$_2$ (2.4 meV).\cite{rf:Broholm87,rf:Broholm91,rf:Mason95} With decreasing temperature, this peak suddenly disappears below $T_{\rm M}=8.3\ {\rm K}$, where the AF phase with nearly fully volume fraction replaces HO. Similar behavior is also observed in pure URu$_2$Si$_2$ under $P$.\cite{rf:Ami2000,rf:Bourdarot2004}
\begin{figure}[tbp]
\begin{center}
\includegraphics[keepaspectratio,width=0.5\textwidth]{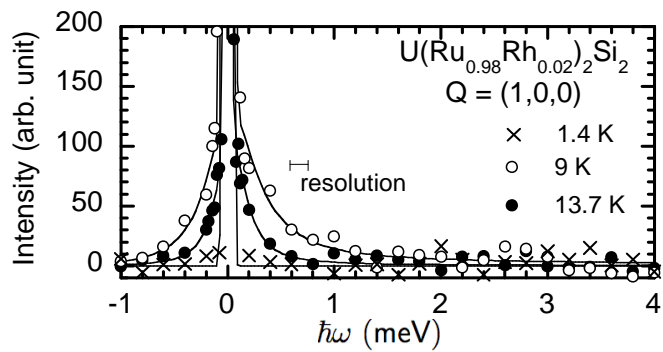}
\end{center}
  \caption{Temperature variations of the inelastic neutron scattering spectra at $Q=(1,0,0)$ for U(Ru$_{0.98}$Rh$_{0.02}$)$_2$Si$_2$. The horizontal bar indicates the width of the energy resolution ($\sim 0.14\ {\rm meV}$) estimated from the FWHM of the incoherent scattering.}
\end{figure}

The disappearance of the magnetic excitations in the AF phase is considered to reflect the features of the low-energy U 5f states in URu$_2$Si$_2$: the matrix elements on the U 5f magnetic moment between the ground state and the low-energy excited states become zero in the AF phase. We suggest that these properties can be explained by assuming the $\Gamma_5$ crystalline-electric-field doublet to be the ground state.\cite{rf:Ami94,rf:Ohkawa99} The $\Gamma_5$ doublet has the degrees of freedom of the dipole moment $J_z$ and the quadrupole moments $J_x^2-J_y^2$ and $J_xJ_y+J_yJ_x$, and they are orthogonal to one another. Therefore, if this doublet splits into the singlets due to the quadrupole ordering, the $J_z$ component can be observed as the magnetic excitation in the inelastic neutron scattering measurement. By switching the order parameter to be $J_z$, on the other hand, the magnetic excitation cannot be detected by means of usual neutron scattering technique, because there are only quadrupole matrix elements between the singlets.\cite{rf:Ohkawa99} This is consistent with our inelastic neutron scattering experiments if we assume the HO parameter to be quadrupole. However, the bulk properties indicate that the itinerant heavy quasi particles play important roles in the evolution of HO. We thus need to take account of both the localized and itinerant features on the HO phase.

\section{Conclusion}
We have discussed the relationship between HO and AF phases in URu$_2$Si$_2$ on the basis of recent microscopic investigations including our neutron scattering experiments under $P$, $\sigma$ and Rh-substitution conditions, and suggest the following features on these two phases: (a) HO and AF are not microscopically coexistent but spatially separated, (b) the competition between HO and AF is governed by the tetragonal $c/a$ ratio, (c) the inhomogeneous development of the AF state is ascribed to the distribution of the $c/a$ ratio in the sample, and (d) the magnetic excitation observed in the HO phase disappears in the AF phase. Various inconsistencies concerning the ``weak" AF moment at ambient pressure are expected to be resolved by taking account of above features. From the feature (d), in addition, we propose the quadrupole order as a most promising candidate for HO. Nevertheless, the nature of HO is not sufficiently elucidated yet for lack of direct observation of HO. Further microscopic experiments will be needed to clear up the issue on HO.

\section*{Acknowledgments}
We are grateful to the people named below for many fruitful discussions and collaboration: S.\ Kawarazaki, J.A.\ Mydosh, Y.\ Miyako, N. Metoki, M. Sato, K.\ Kuwahara, D.\ Andreica, A.\ Amato, A.\ Schenck, T.\ Honma, K.\ Tenya, H.\ Yoshizawa, N.\ Aso, T.\ Sakakibara, S.\ Itoh and I.\ Kawasaki. This work was supported by a Grant-in-Aid for Scientific Research from the Ministry of Education, Culture, Sports, Science and Technology.

\end{document}